\documentclass[twocolumn]{aastex61}

\usepackage[normalem]{ulem}
\begin{document}

\def\liu#1{{\bf #1}}

\shorttitle{The origins of UV-optical color gradients in SFGs at $\lowercase{z}\sim2$}
\title{
The Origins of UV-optical Color Gradients in Star-forming Galaxies at $\lowercase{z}\sim2$: 
Predominant Dust Gradients But Negligible \lowercase{s}SFR Gradients
}
\correspondingauthor{F. S. Liu}
\email{Email: fsliu@synu.edu.cn}

\author{F. S. Liu $^{\color{blue} \dagger}$}
\affil{College of Physical Science and Technology, Shenyang Normal University, Shenyang 110034, China}
\affiliation{University of California Observatories and the Department of Astronomy and Astrophysics,
University of California, Santa Cruz, CA 95064, USA}

\author{Dongfei Jiang}
\affil{College of Physical Science and Technology, Shenyang Normal University, Shenyang 110034, China}
\affiliation{Purple Mountain Observatory, Chinese Academy of Sciences,
2 West-Beijing Road, Nanjing 210008, China}

\author{S. M. Faber}
\affiliation{University of California Observatories and the Department of Astronomy and Astrophysics,
University of California, Santa Cruz, CA 95064, USA}

\author{David C. Koo}
\affiliation{University of California Observatories and the Department of Astronomy and Astrophysics,
University of California, Santa Cruz, CA 95064, USA}

\author{Hassen M. Yesuf}
\affiliation{University of California Observatories and the Department of Astronomy and Astrophysics,
University of California, Santa Cruz, CA 95064, USA}

\author{Sandro Tacchella}
\affiliation{Department of Physics, Institute for Astronomy, ETH Zurich, CH-8093 Zurich, Switzerland}

\author{Shude Mao}
\affiliation{Physics Department and Tsinghua Centre for Astrophysics, Tsinghua University, Beijing 100084, China}
\affiliation{National Astronomical Observatories, Chinese Academy of Sciences, A20 Datun Road, Beijing 100012, China}
\affiliation{Jodrell Bank Centre for Astrophysics, School of Physics and Astronomy, The University of Manchester, Oxford Road, Manchester M13 9PL, UK}

\author{Weichen Wang}
\affiliation{Department of Physics \& Astronomy, Johns Hopkins University, 3400 N. Charles Street, Baltimore, MD 21218, USA}

\author{Yicheng Guo}
\affiliation{University of California Observatories and the Department of Astronomy and Astrophysics,
University of California, Santa Cruz, CA 95064, USA}

\author{Jerome J. Fang}
\affiliation{University of California Observatories and the Department of Astronomy and Astrophysics,
University of California, Santa Cruz, CA 95064, USA}
\affiliation{Orange Coast College, Costa Mesa, CA 92626, USA}

\author{Guillermo Barro}
\affiliation{University of California Observatories and the Department of Astronomy and Astrophysics,
University of California, Santa Cruz, CA 95064, USA}
\affiliation{Department of Astronomy, University of California, Berkeley, CA 94720-3411, USA}

\author{Xianzhong Zheng}
\affiliation{Purple Mountain Observatory, Chinese Academy of Sciences,
2 West-Beijing Road, Nanjing 210008, China}
\affiliation{Chinese Academy of Sciences South America Center for Astronomy,
China-Chile Joint Center for Astronomy, Camino El Observatorio 1515, Las Condes, Santiago, Chile}

\author{Meng Jia}
\affil{College of Physical Science and Technology, Shenyang Normal University, Shenyang 110034, China}

\author{Wei Tong}
\affil{College of Physical Science and Technology, Shenyang Normal University, Shenyang 110034, China}

\author{Lu Liu}
\affil{College of Physical Science and Technology, Shenyang Normal University, Shenyang 110034, China}

\author{Xianmin Meng}
\affil{National Astronomical Observatories, Chinese Academy of Sciences, A20 Datun Road, Beijing 100012, China}


\begin{abstract}

The rest-frame UV-optical (i.e., $NUV-B$) color is sensitive to both 
low-level recent star formation (specific star formation rate - sSFR) and dust.
In this Letter, we extend our previous work on the origins of 
$NUV-B$ color gradients in star-forming galaxies (SFGs) 
at $z\sim1$ to those at $z\sim2$. We use a sample of 1335 
large (semi-major axis radius $R_{\rm SMA}>0.''18$) SFGs with extended 
UV emission out to $2R_{\rm SMA}$ 
in the mass range $M_{\ast} = 10^{9}-10^{11}M_{\odot}$ at $1.5<z<2.8$ in the CANDELS/GOODS-S and UDS fields. 
We show that these SFGs generally have negative $NUV-B$ color gradients (redder centers), 
and their color gradients strongly increase with galaxy mass. 
We also show that the global rest-frame $FUV-NUV$ color is approximately linear with 
$A_{\rm V}$, which is derived by modeling the observed integrated FUV to NIR spectral energy distributions 
of the galaxies. Applying this integrated calibration to 
our spatially-resolved data, we find a negative dust gradient (more dust extinguished in the centers), 
which steadily becomes steeper with galaxy mass. 
We further find that the $NUV-B$ color gradients become nearly zero after correcting for dust gradients 
regardless of galaxy mass. This indicates that the sSFR gradients are negligible 
and dust reddening is likely the principal cause of negative UV-optical 
color gradients in these SFGs. Our findings support that the 
buildup of the stellar mass in SFGs at the Cosmic Noon is self-similar inside $2R_{\rm SMA}$. 


\end{abstract}

\section{Introduction}

Investigating the spatial distribution of star-formation 
is a powerful way to understand how stellar mass is built up and where the star-formation is 
shut down in galaxies as they evolve along the star-forming main sequence (SFMS). 
It has been known that rest-frame UV-optical (i.e., $NUV-B$) color is sensitive to 
both low-level recent star formation (i.e., specific star formation rate - sSFR) 
and dust ($A_{\rm V}$), 
but it is insensitive to the metallicity \citep[][]{Kaviraj07,Pan15}. 
Thus, UV-optical star formation measurements 
are ambiguous without accurate dust correction, especially for 
high-redshift star-forming galaxies (SFGs).

Radial sSFR and dust gradients in distant galaxies 
have not been fully explored to date. There are only few related papers on this topic. 
By stacking the {\it HST} multi-band imaging, 
\citet[][]{Wuyts12} studied the resolved colors and stellar populations
of a few hundred SFGs with $M_{\ast} > 10^{10}M_{\odot}$ at $0.5<z<2.5$. 
They found evidence for redder colors, lower sSFR and increased dust attenuation in the centers of galaxies. 
\citet[][]{Tacchella15} used $\rm H{\alpha}$ fluxes to measure 
the sSFR gradients in $z\sim2.2$ SFGs. 
They claimed rather shallow sSFR gradients at low masses ($M_{\ast} < \sim 10^{11}M_{\odot}$) 
and significant sSFR gradients at $M_{\ast} \sim 10^{11}M_{\odot}$. 
Note that they corrected for dust reddening assuming a flat attenuation profile. 
In a series of papers, \citet{Nelson12,Nelson16b,Nelson16a} studied the maps of sSFR traced by $\rm H{\alpha}$ 
and of dust in SFGs at moderate redshifts ($z\sim1$ and $z\sim1.4$) 
by stacking the spatially-resolved spectra of 3D-HST.
\citet[][]{Nelson12} showed that the $\rm H{\alpha}$ sizes of massive galaxies
are bigger than their rest-frame $R$-band sizes. 
\citet[][]{Nelson16b} showed that the $EW({\rm H{\alpha}})$ is flat with radius for
the low-mass ($M_{\ast} = 10^{9}-10^{9.5}M_{\odot}$) galaxies, 
while it falls by a factor 
of $\sim2$ on average from the center to twice the effective radius for more massive galaxies.
These findings suggested that massive SFGs at moderate redshifts buildup 
their stellar masses from the inside out,
while the low-mass SFGs grow in a self-similar way irrespective of the radial distance.
Note that dust correction was not done in these two works. 
Later, \citet[][]{Nelson16a} corrected the data in their previous papers for dust by 
using the Balmer decrement ($\rm H{\alpha}/H{\beta}$). 
As a result, central dust was found to be a huge factor in establishing radial colors and sSFR 
gradients in galaxies with a mean mass of $\langle M_{\ast} \rangle \sim 10^{10.2}M_{\odot}$. Galaxies 
with $\langle M_{\ast} \rangle \sim 10^{9.2}M_{\odot}$ have little dust attenuation at all radii. 

Recently, \citet[][hereafter Paper I]{Liu16} used high-resolution {\it HST} optical-IR 
imaging (in observed $B$ through $H$ band) to measure the rest-frame $NUV-B$ color gradients 
in the main-sequence SFGs at $z\sim1$. In Paper I, after correcting for dust reddening, 
the radial $NUV-B$ color gradients were shown to be nearly zero in SFGs 
with $M_{\ast} < 10^{10}M_{\odot}$, but 
significant residual color gradients were found in SFGs with $M_{\ast} > 10^{10.5}M_{\odot}$. 
Dust gradients were determined by fitting reddened stellar population models to 
the spatially-resolved spectral energy distributions (SEDs) with {\tt FAST} \citep[][]{Kriek+09}.
These findings implied that at $z\sim1$ dust reddening is the principal cause of rest-frame $NUV-B$ 
color gradients in low-mass SFGs, while for high-mass SFGs, 
age gradients are also an important factor. 
More recently, \citet[][]{Wang17} expanded our Paper I and 
re-visited the dust and sSFR gradients in $z=0.4-1.4$ SFGs on the main-sequence 
inferred from the $UVI$ ($U-V$ versus $V-I$) diagrams. Their conclusion is 
generally consistent with that of Paper I, except that their estimated central sSFR for massive SFGs 
are 2-3 times smaller than ours.
In this work, we extend our Paper I to $z\sim2$ with a 
sample of 1335 large SFGs with extended UV emission out 
to twice the semi-major axis radius in the mass range $M_{\ast} = 10^{9}-10^{11}M_{\odot}$ 
at $1.5<z<2.8$ in the CANDELS/GOODS-S and UDS fields. 
We show that the global rest-frame $FUV-NUV$ color is approximately linear 
with $A_{\rm V}$ and the dust-corrected $NUV-B$ color is indeed a good 
tracer of sSFR. Applying these calibrations to our spatially-resolved data, 
we examine the effects of dust gradients and sSFR gradients on 
the $NUV-B$ color gradients in these SFGs at the Cosmic Noon 
and further discuss their link to the stellar mass assembly.  

Throughout the Letter, we adopt a cosmology with a matter density 
parameter $\Omega_{\rm m}=0.3$, a cosmological constant 
$\Omega_{\rm \Lambda}=0.7$ and a Hubble constant of ${\rm H}_{\rm 
0}=70\,{\rm km \, s^{-1} Mpc^{-1}}$. All magnitudes are in the AB system.

\section{DATA}

We select galaxies in our sample from the first two
publicly-available fields of the CANDELS. Namely,
the Great Observatories Origins Deep Survey \citep[GOODS-S,][]{Guo+13}
and the UKIDSS Ultra-Deep Survey \citep[UDS,][]{Galametz13}.
The CANDELS team has made a multi-wavelength catalog based on source
detection in the F160W ($H$) band for each field,
combining the newly obtained CANDELS {\it HST}/WFC3 data with existing public
ground-based and space-based data.
{\it HST} photometry was measured by running {\tt SExtractor}
\citep[][]{Bertin96}
on the point spread function (PSF)-matched images in the dual-image mode,
with the F160W image as the detection image.
Photometry in ground-based and IRAC images,
whose resolutions are much lower than that of the F160W images,
was measured with {\tt TFIT} \citep[][]{Laidler07},
which fit the PSF-smoothed high-resolution image templates to the low-resolution images
to measure the fluxes in the low-resolution images. We refer readers to
\citet[][]{Guo+13} and \citet[][]{Galametz13} for details
on these data and the reduction procedure.

Photometric redshifts were estimated from a variety of different codes available
in the literature,
which are then combined to improve the individual performance
\citep[][]{Dahlen13}.
Rest-frame total magnitudes in various standard filters, from $FUV$ to $K$,
were computed using the best available redshifts (spectroscopic or photometric) and multi-wavelength
photometry using {\tt EAZY} \citep[][]{Brammer+08}, which fits a set of galaxy SED 
templates to the observed photometry.
Stellar masses come from the CANDELS official catalog released by
\citet[][]{Santini+15}, which combine the results from ten separate SED fitting methods.
A \citet[][]{Chabrier03} initial mass function (IMF) is assumed.
%
Semi-major axis radius ($R_{\rm SMA}$) and axis ratio ($b/a$) were measured 
from the {\it HST}/WFC3 F160W images using {\tt GALFIT} \citep[][]{Peng+02} 
by \citet[][]{vdWel+12}. 
SFRs come from rest-frame $NUV$ luminosities at $\lambda\approx2800\AA$ 
that are corrected for extinction by assuming a Calzetti law ($\rm A_{2800}\approx1.79A_V$):
$ SFR_{\rm NUV,cor}[M_{\odot}{yr}^{-1}]=2.59\times10^{-10}L_{\rm NUV,cor}[L_{\odot}]$ 
\citep[][]{Kennicutt12}. 
We prefer this approach because of its simplicity and more direct relation to
the observed SED. The rates of star-formation derived from a combination of unobscured UV 
and IR emission ($SFR_{\rm UV+IR}$) are nominally a more faithful measure than $SFR_{\rm NUV,cor}$, 
since it incorporates a $direct$ measure of obscured star formation. However, 
$L_{\rm IR}$ is usually overestimated in galaxies above $z\sim1.5$, 
where observed 24 $\mu$m probes PAH emission \citep[][]{Tielens08}. 
Fitting local IR templates will return a value of $L_{\rm IR}$ that is systematically 
too high \citep[][]{Salim09}. Given the theoretical uncertainties still surrounding 
the origin of warm dust in galaxies, it is reasonable to adopt the NUV-NIR SED-fitting rates, 
as we do here, pending further developments in our understanding of mid-IR galaxy SEDs. 
The median $A_{\rm V}$ was computed by combining results from four methods (labeled 
$2a_{\tau}$, $12a$, $13a_{\tau}$ and $14a$) in \citet[][]{Santini+15}. 
The four methods were chosen with the same assumptions (Chabrier IMF and Calzetti extinction law). 

The {\it HST} based multi-wavelength and multi-aperture photometry catalogs 
with improved local background subtraction were built for galaxies in the CANDELS fields 
(Liu et al. in prep.), which include the radial profiles of 
observed surface brightness and cumulative magnitude 
in the {\it HST}/WFC3 (F105W, F125W F140W, and F160W) bands and {\it HST}/ACS 
(F435W, F606W, F775W, F814W and F850LP) bands if available. 
The photometry were performed on the multi-band, PSF-matched images. 

\section{Sample Selection}

The full GOODS-S and UDS catalogs contain 34,930 and 35,932 objects, respectively. 
The parent sample used in this work is constructed 
by applying the following cuts to the above data: \\

1. Observed F160W($H$) magnitude $ H < 24.5$ and the {\tt GALFIT} quality $\rm flag = 0$ in F160W 
\citep[][]{vdWel+12} to ensure well-constrained {\tt GALFIT} measurements 
and eliminate doubles, mergers, and disturbed objects. \\

2. Photometry quality flag $\rm PhotFlag = 0$ to exclude spurious sources. \\

3. SExtractor $\rm CLASS\_STAR < 0.9$ to reduce contamination by stars. \\

4. Redshifts within $1.5<z<2.8$ for GOODS-S and $2.2<z<2.8$ for UDS to roughly cover 
rest-frame $FUV$ to $B$ simultaneously. Note that the shortest observed band in GOODS-S available 
is F435W($B$) and the shortest one in UDS is F606W($V$). \\ 

5. Stellar masses at $\rm 10^{9} < M_{\ast}/M_{\odot} < 10^{11}$ to maintain 
$\sim90\%$ ($\sim75\%$) completeness limit at $z=1.5$ ($z=2.8$) 
for SFGs \citep[][]{vdWel2014,Morishita15}. \\

6. $R_{\rm SMA}>0.18^{\prime\prime}~(\rm 3~pixels)$ to minimize the effect of
PSF-matching on color gradient measurement. \\

7. Well-constrained measurements of surface brightness profiles 
from center to $2R_{\rm SMA}$ in observed F435W($B$) for $1.5<z<2.2$ in GOODS-S and 
ones in observed F606W($V$) for $2.2<z<2.8$ in both GOODS-S and UDS to 
guarantee sample galaxies with extended rest-frame UV emission.\\   

After the cuts 1-5, we obtain 2388 galaxies: 
1666 from GOODS-S and 722 from UDS. After the cuts 1-7, we 
obtain 1430 galaxies in total. 
We then utilize the rest-frame $UVJ$ ($U-V$ versus $V-J$) diagram 
\citep[$(U - V) < 0.88 \times (V - J) + 0.49$,][]{Williams09} 
to select 1405 SFGs (see the left panel in Figure 1). 
Furthermore, we follow the method used in Paper I to select SFGs 
near the ridge-line of the SFMS (see the right panel in Figure 1). 
After excluding 70 transition galaxies, which are defined as 
galaxies with offsets from the best-fit main-sequence relation 
($log~sSFR_{\rm NUV,cor}/yr^{-1}=-0.25\pm0.01 {\times} logM_{\ast}/M_{\odot}-6.20\pm0.13$) 
of ${\rm \Delta}~log~sSFR_{\rm NUV,cor}<{\rm -0.45~dex}$ (below the dashed line), we 
focus on 1335 main-sequence SFGs.
We note that the majority of very dusty SFGs are excluded 
by the selection cuts 6 and 7 (see the left panel in Figure 1), 
which are not probed in this analysis.

\section{Results and Analysis}

We investigate the global properties of the sample galaxies to find an accessible and good 
indicator of dust attenuation ($A_{\rm V}$) that can be used in our spatially-resolved analysis 
and check whether rest-frame $NUV-B$ color is a good tracer of sSFR after removing 
dust effect. One of the most common methods for dust determination at high redshift is 
fitting reddened stellar population models to the integrated broad-band 
SEDs of galaxies \citep[e.g.,][]{Kriek+09}. 
The long-wavelength (i.e., rest-frame $J$) data is usually needed by this method. 
It has been shown that this method 
is closely related to the $UVJ$ method that distinguishes dust reddening 
from old stars \citep[][]{Patel11}. 
Unfortunately, the high-resolution {\it HST} imaging in CANDELS ends at observed $H$ band, 
which roughly corresponds to a cut at rest-frame $B$ to $V$ for our galaxies. 
The lack of long-wavelength high-resolution data can not meet the need of distinguishing dust 
reddening from age in spatially-resolved analysis. 
Nevertheless, the rest-frame $FUV-NUV$ color is accessible in both integrated and resolved data, 
which has been widely applied to evaluate the $UV$ slope $\beta$ and thus determine dust attenuation 
\citep[e.g.,][]{Buat05,Munoz07,Reddy12}. 
In the left panel of Figure 2, we show the relation of global $A_{\rm V}$ versus 
rest-frame $FUV-NUV$ for our SFGs. It can be seen that rest-frame $FUV-NUV$ color 
is approximately linear with $A_{\rm V}$. 
The best linear fit is given as $ A_{\rm V} = 1.38\pm0.02 \times (FUV-NUV) + 0.30\pm0.01$. 
In the right panel of Figure 2, we show the relation of $log~sSFR_{\rm NUV,cor}$ versus $(NUV-B)_{\rm dc}$. 
The dust correction exploits the median $A_{\rm V}$ derived 
by modeling the observed integrated FUV to NIR SEDs and assumes a Calzetti law. 
This plot demonstrates that the dust-corrected $NUV-B$ color is indeed a good 
tracer of sSFR for our SFGs. 
The best polynomial fit to this relation is given as 
$log~sSFR_{\rm NUV,cor}/yr^{-1}=0.62\pm0.01 \times {(NUV-B)_{\rm dc}}^2-1.78\pm0.02 \times 
(NUV-B)_{\rm dc} -7.88\pm0.13$. 

Applying the above calibrations from integrated photometry and SED modeling 
to our spatially-resolved data, we can infer the radial $A_{\rm V}$ gradients and 
sSFR gradients to disentangle their effects on the $NUV-B$ color gradients. 
For this work we computed the rest-frame $FUV$, $NUV$ and $B$ band surface brightness profiles of each 
sample galaxy using {\tt EAZY} \citep[][]{Brammer+08} as well (refer to Figure 2 in Paper I). 
In Figure 3, we show the raw dust-reddened $FUV-NUV$ profiles and 
inferred $A_{\rm V}$ profiles, which are normalised by their $\rm R_{SMA}$ in 
arcsec (upper) and are shown in physical radius (lower), respecitvely.
The individual galaxy profiles are shown with gray lines. 
To quantify the general trends, we used the linear model {\tt lm} function  
in {\tt R} programming language to fit a straight line as a mean of
the individual profiles between PSF FWHM ($0.18^{\prime\prime}$) and $2R_{\rm SMA}$ in each panel. 
The best-fit slopes and intercepts are presented in Table 1. 
The best-fit models with $2\sigma$ lower and upper limits are shown as shaded regions 
in Figure 3.
It is observed that these SFGs generally 
have negative $FUV-NUV$ color gradients (redder centers) and thus have negative dust gradients 
(more dust extinguished in the centers). A steady increase of negative dust gradient 
(the slope tends to become steeper) with galaxy mass is also observed. 

In Figure 4, we show the profiles of raw dust-reddened $NUV-B$, 
dust-corrected $NUV-B$ and inferred sSFR in each adopted mass bin, respectively. 
The shade regions show the best-fit linear models with $2\sigma$ lower and upper limits 
to all individual profiles. The best-fit slopes and intercepts are also presented 
in Table 1. 
The dust correction exploits inferred $A_{\rm V}$ profiles shown in the bottom panels of Figure 3 
and assumes a Calzetti law.
sSFR profiles were computed with the $(NUV-B)_{\rm dc}$ profiles 
after applying the calibration shown in the right panel of Figure 2. 
It can be seen that these SFGs generally have negative $NUV-B$ 
color gradients (redder centers), and the color gradients strongly increase 
with galaxy mass. However, after correcting for dust reddening, 
the $NUV-B$ color profiles become nearly flat (the slopes are within $\pm0.05$), 
which results in nearly flat sSFR gradients. 

To evaluate the PSF effect, the individual rest-frame $FUV$, $NUV$ and $B$ 
band surface brightness profiles in each mass bin are stacked together (taking median 
values) based on the angular distance in arcsec. We then fit the stacking 
surface brightness profiles in each band with a single S\'ersic model convolving 
with CANDELS PSF in F160W. The resulting profiles based on 
PSF-deconvolved S\'ersic models in each bin are shown with green dashed lines in 
Figure 3 and Figure 4, respectively. The slopes and intercepts of the best linear fits to 
these profiles are listed in Table 1 as well.  
The same conclusions can be drawn from these PSF-deconvolved data.

\section{Discussion and Conclusion}

In this Letter, we extend our previous work on the origins of UV-optical 
color gradients in SFGs at $z\sim1$ \citep[][]{Liu16} to those at $z\sim2$, 
using a sample of 1335 main-sequence SFGs with 
extended UV emission in the mass range $M_{\ast} = 10^{9}-10^{11}M_{\odot}$ 
at $1.5<z<2.8$ in the CANDELS/GOODS-S and UDS fields. 
By fitting reddened stellar population models to 
the integrated SEDs from observed FUV to NIR, 
we calibrate $A_{\rm V}$ with the rest-frame $FUV-NUV$ for these SFGs. 
We demonstrate that rest-frame $NUV-B$ color is indeed a good 
tracer of sSFR after correcting for dust reddening. 
Applying these calibrations to our spatially-resolved data that ends at observed $H$, 
we infer the radial $A_{\rm V}$ gradients and sSFR gradients and demonstrate 
their effects on the $NUV-B$ color gradients. 
We find a steady increase of negative dust gradient with galaxy mass. The SFGs generally have negative $NUV-B$ 
color gradients, and the color gradients strongly increase with galaxy mass.
After correcting for dust gradients, the $NUV-B$ color profiles become nearly flat 
regardless of galaxy mass, which indicates that the sSFR gradients are negligibly small. 
These findings imply that dust reddening is likely the principal cause of negative UV-optical 
color gradients in these SFGs. The findings support that 
at $z\sim2$ the SFGs buildup their stellar masses in a self-similar way. 


Here we compare our results to those of \citet[][]{Tacchella15,Tacchella17}, 
who explored sSFR gradients in $z\sim2.2$ galaxies for a small sample. 
In \citet[][]{Tacchella15}, 
they corrected for dust reddening assuming a flat attenuation profile and 
claimed rather shallow sSFR gradients in low-mass ($M_{\ast} < 10^{11}M_{\odot}$) galaxies 
but significant sSFR gradients in galaxies with $M_{\ast} \sim 10^{11}M_{\odot}$. 
In \citet[][]{Tacchella17}, 
they used the $UV$-$\beta$ technique to correct for dust reddening 
and found flat sSFR profiles for $z\sim2.2$ galaxies in the mass range $M_{\ast} = 10^{10}-10^{11}M_{\odot}$, 
which is consistent with our finding. 
Our finding is also consistent with that from cosmological zoom-in simulations \citep[][]{Tacchella16}.

If the sSFR profiles appear to be constant with radius 
at all radii and all times, this would yield SFGs with constant stellar-mass effective radii. 
In contrast, light-weighted effective radii are seen to 
increase roughly as $M_{\star}^{0.3}$ \citep[e.g.,][]{Patel13,vDokkum13}. 
It has been known that mass-weighted radii are smaller than 
light-weighted for SFGs \citep[][]{Szomoru13}. 
The discrepancy can also be reconciled if galaxies grow mainly outside $\rm 2R_{SMA}$ 
via star formation \citep[][]{Tacchella16} or minor mergers 
\citep[][]{Welker17}. We leave this problem open because 
our data only sample the regions inside $\rm 2R_{SMA}$.

We note that the {\it HST} drizzled WFC3 images have the spatial resolution of 
$\rm FWHM \sim 0.18^{\prime\prime}$ (3 pixels). Therefore, the color gradients 
in this very central region are missed by this analysis.
This is the best that can be done with the present imaging data available.
We stress that major conclusions in this paper depend on the SED modeling 
assumptions applied to the CANDELS data. 
The majority of our assumptions are single-$\tau$ solar metallicity models and 
the dust extinction law is assumed to be the Calzetti law.
\citet[][]{Wang17} showed that the dust-reddened radial color variance for the main-sequence SFGs 
at $z\sim1$ run almost parallel to the Calzetti vectors in the $UVI$ diagram. 
But whether the extinction curve of SFGs at $z\sim2$ also follows the Calzetti law is still unknown. 
These assumptions are standard and have been used in all of high-$z$ studies. 
This paper does not attempt to justify these current state of the art assumptions, but 
takes the standard assumptions as given and aims to see where they lead to. 
Future works should investigate the consequences 
of more realistic stellar population models, metallicity, and extinction law.   

\acknowledgments

We acknowledge the anonymous referee for a constructive report that significantly
improved this paper. 
This work was supported by the National Science Foundation of China (Grant No. 11573017 to FL 
and No. 11333003, 11390372 to SM). XZ is supported by the National Basic Research Program of 
China (973 Program 2013CB834900) and the Chinese Academy of Sciences (CAS) through a grant 
to the CAS South America Center for Astronomy (CASSACA) in Santiago, Chile.

\nocite{*}

\begin{figure*}
\centering
\includegraphics[angle=0,width=1.0\textwidth]{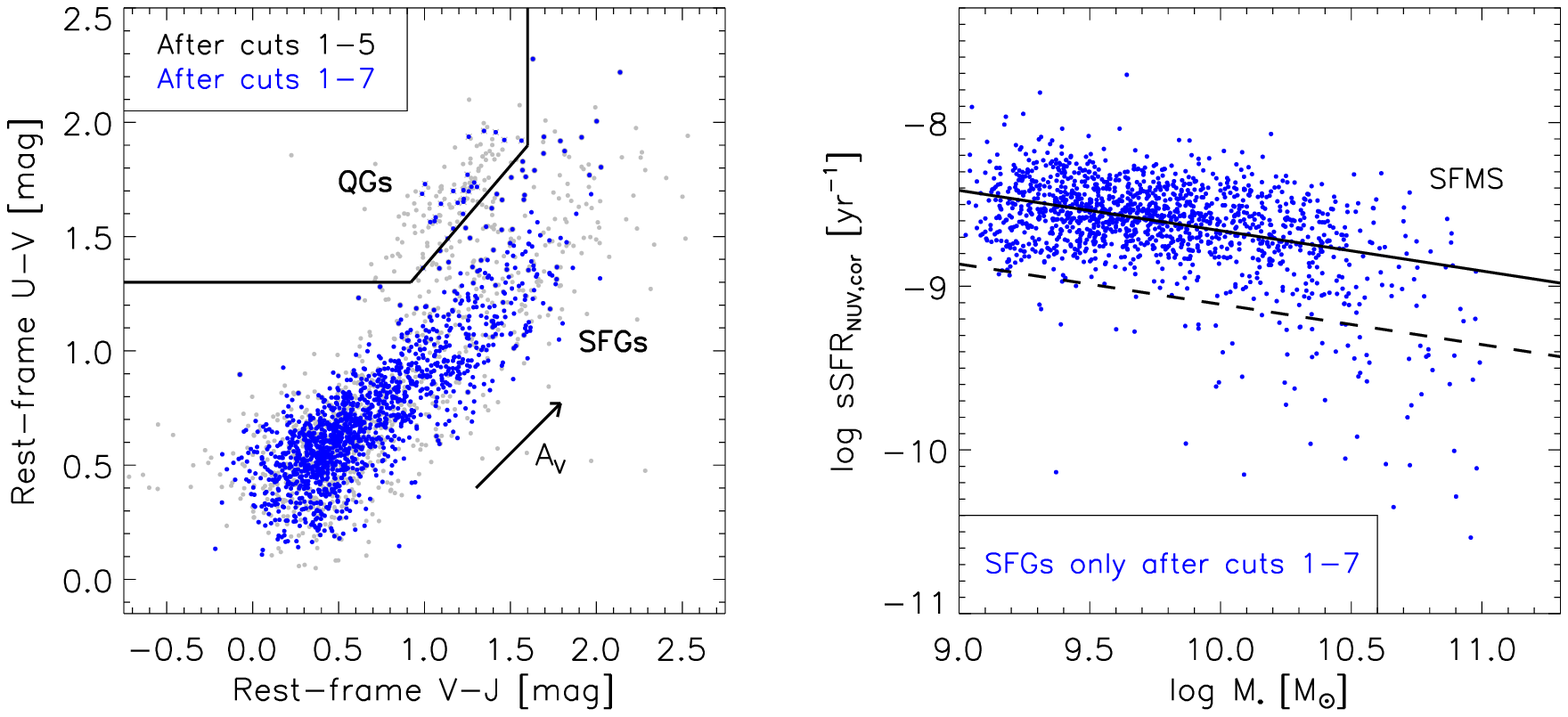}
\caption{
Left panel: Rest-frame global $UVJ$ diagram for the parent samples
after applying the selection cuts 1-5 and 1-7 (see \S1), respectively.
The solid lines indicate the classification criterion provided
by \citet[][]{Williams09}. The arrow shows the Calzetti vector.
Right panel: sSFR vs. stellar mass
for only $UVJ$-defined SFGs after the cuts 1-7.
The solid line shows the best-fit linear relation to the SFMS. The transition galaxies,
defined to have residuals ${\rm \Delta}~log~sSFR_{\rm NUV,cor}<{\rm -0.45~dex}$ (below the dashed line),
are excluded in this analysis.
\label{sample}}
\end{figure*}

\begin{figure*}
\centering
\includegraphics[angle=0,width=1.0\textwidth]{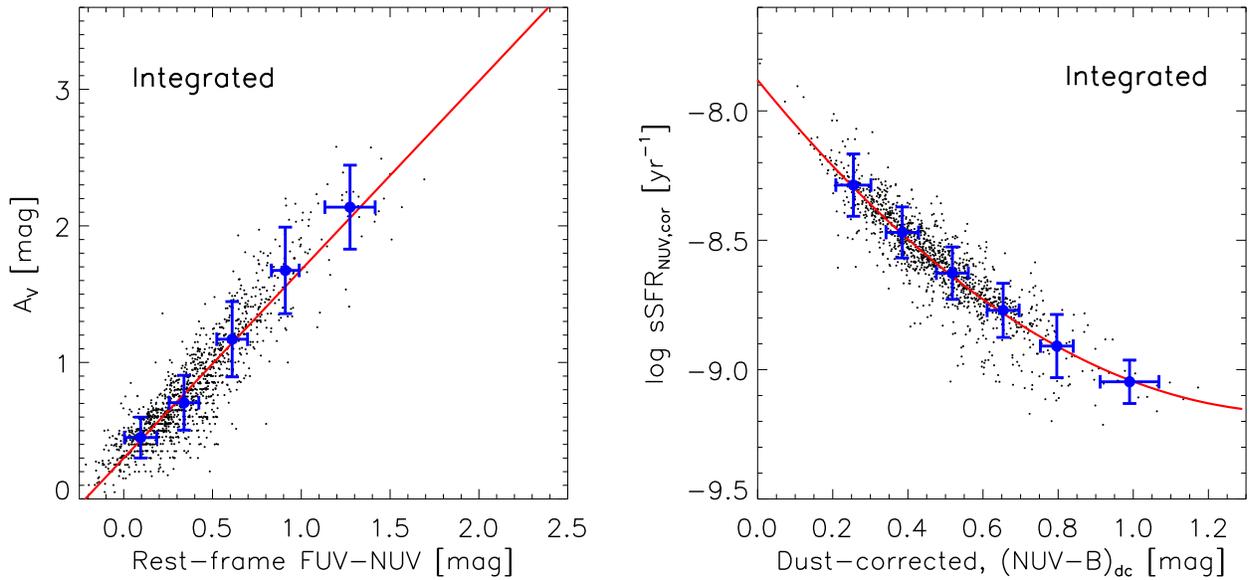}
\caption{
Correlations of $A_V$ versus $FUV-NUV$ (left) and $ log~sSFR_{\rm NUV,cor}$
versus $(NUV-B)_{\rm dc}$ (right) for global (integrated) SFGs.
The red line in the left panel is the best linear fit to the relation.
The red line in the right panel is the best polynomial fit to the relation.
The two plots will be used as calibrations to infer dust and sSSFR radial profiles
of the galaxies in this study.
\label{sample_uvj}}
\end{figure*}

\begin{figure*}
\centering
\includegraphics[angle=0,width=1.0\textwidth]{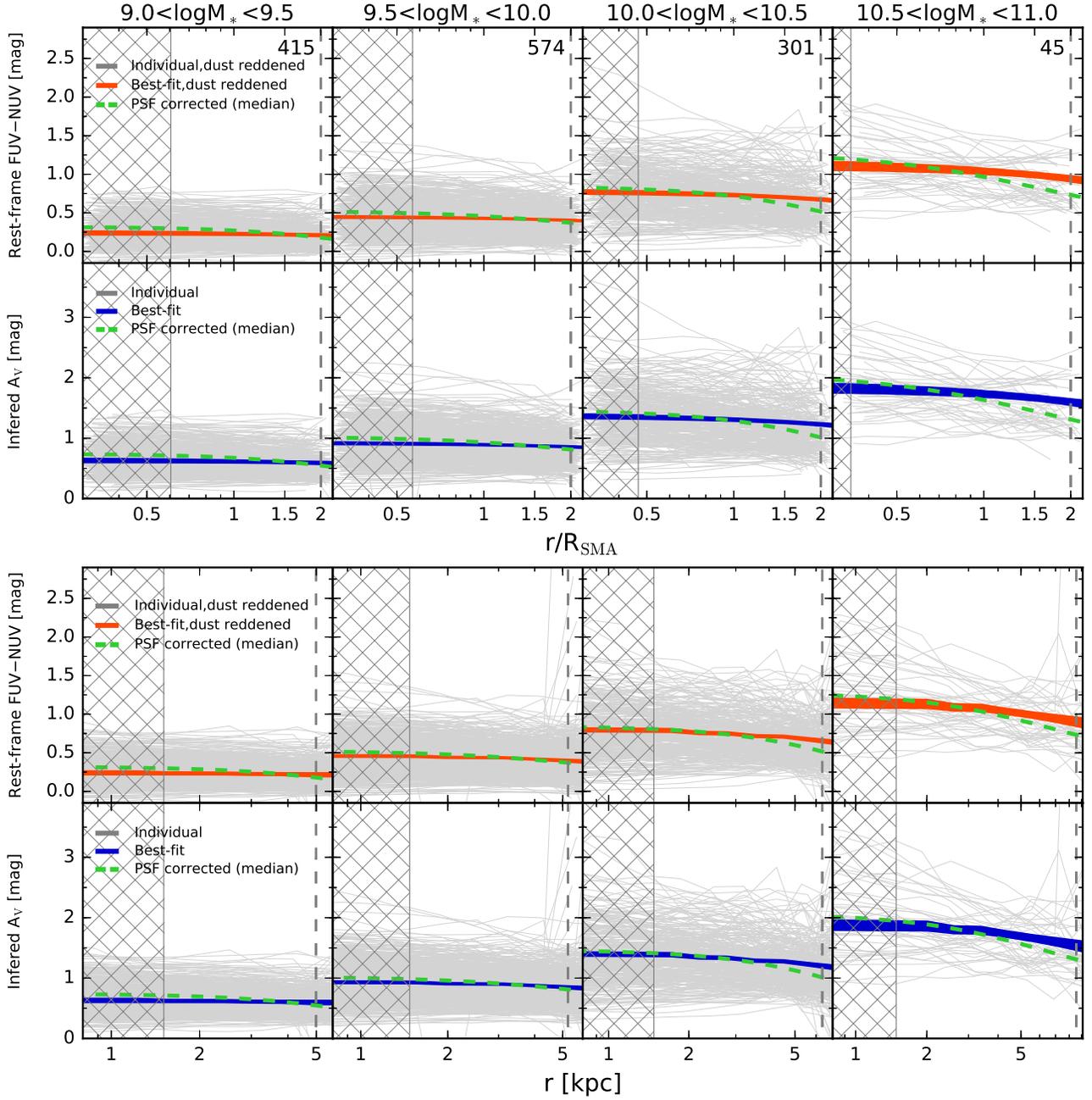}
\caption{
Rest-frame dust-reddened $FUV-NUV$ profiles and inferred $A_V$ profiles in each mass
bin, which are normalised by their $\rm R_{SMA}$ in arcsec (upper) and
are shown in physical radius (lower), respecitvely.
The individual profiles are shown with gray lines.
The shade regions show the best-fit linear models with $2\sigma$ lower and upper limits
to all individual profiles (no PSF-correction). The green dashed lines are PSF-corrected
median profiles. The gray grids indicate the regions within
the median values of twice the PSF radius ($0.18^{\prime\prime}$).
The vertical dashed lines indicate the positions of $2R_{\rm SMA}$ (median values for
the profiles in kpc).
\label{sample_uvj}}
\end{figure*}

\begin{figure*}
\centering
\includegraphics[angle=0,width=1.0\textwidth]{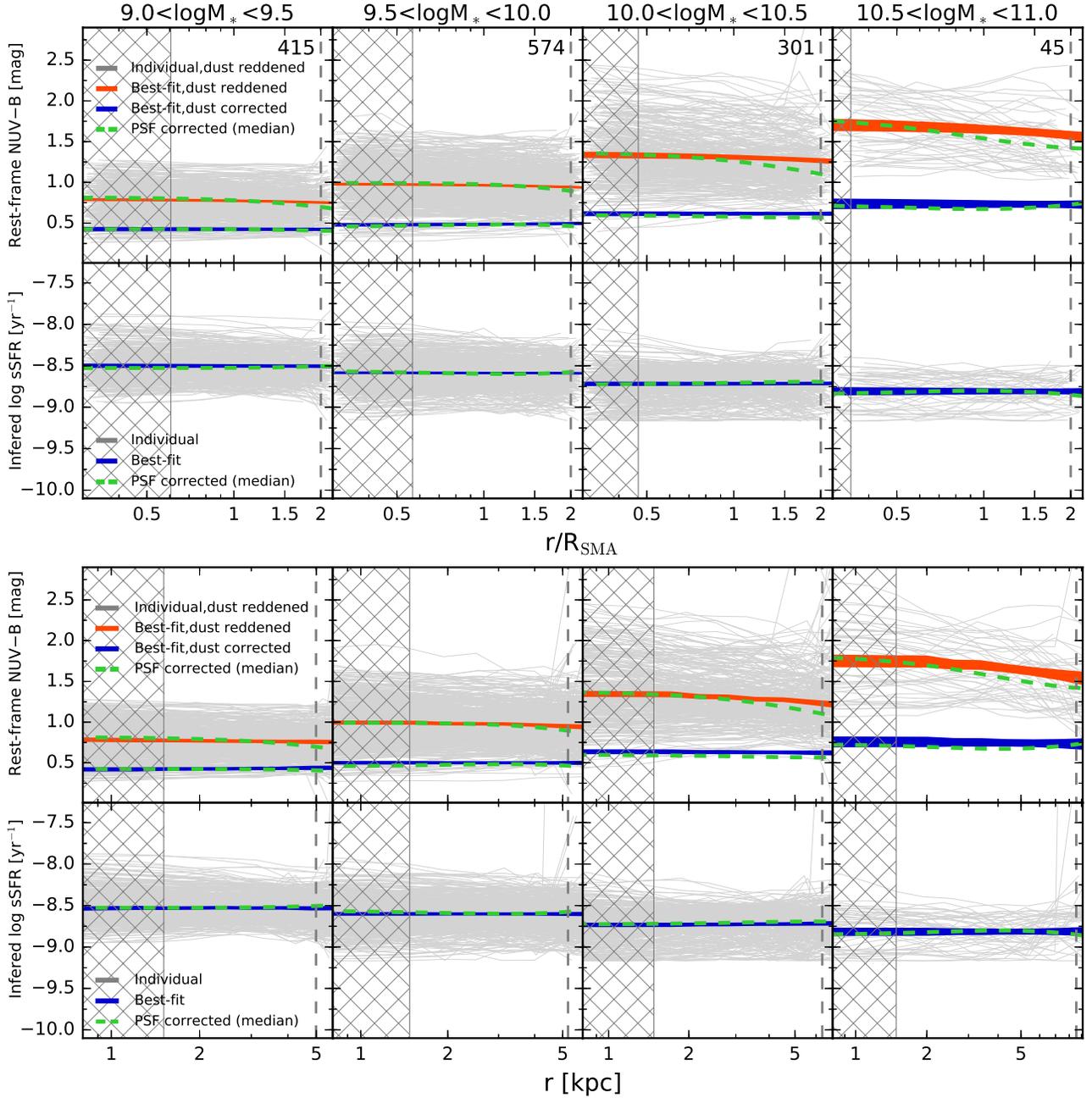}
\caption{
Rest-frame $NUV-B$ profiles (both dust-reddened and dust-corrected)
and inferred sSFR profiles in each mass bin,
which are normalised by their $\rm R_{SMA}$ in arcsec (upper) and
are shown in physical radius (lower), respecitvely.
Only the individual galaxy profiles are shown with gray lines for
dust-reddened $NUV-B$ and inferred sSFR.
The shade regions show the best-fit linear models with $2\sigma$ lower and upper limits
to all individual profiles (no PSF-correction). The green dashed lines are PSF-corrected
median profiles. The gray grids indicate the regions within the median values of 
twice the PSF radius ($0.18^{\prime\prime}$).
The vertical dashed lines indicate the positions of $2R_{\rm SMA}$ (median values for
the profiles in kpc).
\label{sample_uvj}}
\end{figure*}


\begin{deluxetable*}{ccccccccccc}[htb!]
\tablecaption{Parameters of the best-fit linear models to our profiles}
\tablenum{1}
\tabletypesize{\tiny}
\tablehead{
\colhead{} & \colhead{} &
\multicolumn2c{$\rm 9.0<logM_*<9.5$} & \multicolumn2c{$\rm 9.5<logM_*<10.0$} &
\multicolumn2c{$\rm 10.0<logM_*<10.5$} & \multicolumn2c{$\rm 10.5<logM_*<11.0$} \\
\cline{3-4}
\cline{7-8}
\colhead{} & \colhead{Contents} &
\colhead{Slope} & \colhead{Intercept} &
\colhead{Slope} & \colhead{Intercept} &
\colhead{Slope} & \colhead{Intercept} &
\colhead{Slope} & \colhead{Intercept} &
}
\startdata
\hline
\multicolumn8r{Profiles scaled by $\rm R_{SMA}$ in arcsec (no PSF-correction)}\\
\cline{3-10}
{} & {FUV$-$NUV} & -0.104$\pm$0.045 &  0.223$\pm$0.007 & -0.141$\pm$0.025 &  0.420$\pm$0.004 & -0.273$\pm$0.040 &  0.697$\pm$0.007 & -0.390$\pm$0.074 &  0.960$\pm$0.017 \\
{} & {A$\tiny \rm_V$} & -0.143$\pm$0.062 &  0.607$\pm$0.010 & -0.195$\pm$0.035 &  0.879$\pm$0.006 & -0.377$\pm$0.055 &  1.261$\pm$0.010 & -0.538$\pm$0.102 &  1.624$\pm$0.024 \\
{} & {NUV$-$B} & -0.069$\pm$0.006 &  0.753$\pm$0.019 & -0.071$\pm$0.005 &  0.942$\pm$0.001 & -0.199$\pm$0.045 &  1.284$\pm$0.008 & -0.303$\pm$0.087 &  1.590$\pm$0.020 \\
{} & {(NUV$-$B$)\rm\tiny_{dc}$} & -0.008$\pm$0.029 &  0.424$\pm$0.005 &  0.049$\pm$0.019 &  0.488$\pm$0.003 &  0.000$\pm$0.030 &  0.615$\pm$0.005 & -0.018$\pm$0.072 &  0.730$\pm$0.017 \\
{} & {log sSFR} & -0.017$\pm$0.029 & -8.503$\pm$0.004 & -0.012$\pm$0.004 & -8.590$\pm$0.001 &  0.013$\pm$0.025 & -8.714$\pm$0.005 & -0.004$\pm$0.052 & -8.808$\pm$0.012 \\
\cline{3-10}
\multicolumn7r{Profiles in kpc (no PSF-correction)}\\
\cline{3-10}
{} & {FUV$-$NUV} & -0.073$\pm$0.052 &  0.261$\pm$0.024 & -0.167$\pm$0.029 &  0.503$\pm$0.013 & -0.350$\pm$0.040 &  0.891$\pm$0.021 & -0.447$\pm$0.085 &  1.263$\pm$0.049 \\
{} & {A$\tiny \rm_V$} & -0.101$\pm$0.071 &  0.659$\pm$0.034 & -0.231$\pm$0.040 &  0.993$\pm$0.018 & -0.482$\pm$0.055 &  1.528$\pm$0.029 & -0.617$\pm$0.117 &  2.041$\pm$0.068 \\
{} & {NUV$-$B} &  -0.003$\pm$0.036 &  0.753$\pm$0.017 & -0.126$\pm$0.031 &  1.028$\pm$0.014 & -0.286$\pm$0.047 &  1.454$\pm$0.024 & -0.374$\pm$0.096 &  1.852$\pm$0.056 \\
{} & {(NUV$-$B$)\rm\tiny_{dc}$} &  0.046$\pm$0.034 &  0.403$\pm$0.016 & -0.004$\pm$0.022 &  0.502$\pm$0.010 & -0.031$\pm$0.033 &  0.644$\pm$0.017 & -0.048$\pm$0.080 &  0.770$\pm$0.047 \\
{} & {log sSFR} & -0.044$\pm$0.033 & -8.466$\pm$0.016 & -0.004$\pm$0.023 & -8.600$\pm$0.011 &  0.042$\pm$0.027 & -8.745$\pm$0.014 &  0.007$\pm$0.056 & -8.818$\pm$0.033 \\
\hline
\multicolumn8r{PSF-corrected stacking profiles scaled by $\rm R_{SMA}$ in arcsec}\\
\cline{3-10}
{} & {FUV$-$NUV} & -0.214$\pm$0.021 &  0.263$\pm$0.004 & -0.252$\pm$0.013 &  0.454$\pm$0.002 & -0.446$\pm$0.037 &  0.693$\pm$0.007 & -0.686$\pm$0.019 &  0.954$\pm$0.004 \\
{} & {A$\tiny \rm_V$} & -0.296$\pm$0.029 &  0.662$\pm$0.005 & -0.348$\pm$0.017 &  0.926$\pm$0.003 & -0.616$\pm$0.051 &  1.256$\pm$0.010 & -0.946$\pm$0.026 &  1.615$\pm$0.005 \\
{} & {NUV$-$B} & -0.184$\pm$0.021 &  0.772$\pm$0.003 & -0.197$\pm$0.024 &  0.971$\pm$0.004 & -0.373$\pm$0.027 &  1.243$\pm$0.005 & -0.457$\pm$0.009 &  1.543$\pm$0.002 \\
{} & {(NUV$-$B$)\rm\tiny_{dc}$} & -0.028$\pm$0.005 &  0.421$\pm$0.001 & -0.013$\pm$0.015 &  0.480$\pm$0.002 & -0.046$\pm$0.002 &  0.577$\pm$0.000 &  0.045$\pm$0.020 &  0.688$\pm$0.004 \\
{} & {log sSFR} &  0.035$\pm$0.006 & -8.522$\pm$0.001 &  0.015$\pm$0.018 & -8.594$\pm$0.003 &  0.050$\pm$0.002 & -8.705$\pm$0.000 & -0.041$\pm$0.019 & -8.815$\pm$0.004 \\
\cline{3-10}
\multicolumn7r{PSF-corrected stacking profiles in kpc}\\
\cline{3-10}
{} & {FUV$-$NUV} & -0.214$\pm$0.021 &  0.348$\pm$0.010 & -0.223$\pm$0.015 &  0.543$\pm$0.008 & -0.511$\pm$0.034 &  0.957$\pm$0.020 & -0.686$\pm$0.019 &  1.387$\pm$0.013 \\
{} & {A$\tiny \rm_V$} & -0.296$\pm$0.029 &  0.780$\pm$0.014 & -0.308$\pm$0.021 &  1.048$\pm$0.010 & -0.705$\pm$0.047 &  1.620$\pm$0.028 & -0.946$\pm$0.026 &  2.213$\pm$0.018 \\
{} & {NUV$-$B} & -0.184$\pm$0.021 &  0.845$\pm$0.010 & -0.187$\pm$0.023 &  1.030$\pm$0.012 & -0.422$\pm$0.024 &  1.460$\pm$0.014 & -0.457$\pm$0.009 &  1.832$\pm$0.006 \\
{} & {(NUV$-$B$)\rm\tiny_{dc}$} & -0.028$\pm$0.005 &  0.432$\pm$0.002 &  0.006$\pm$0.013 &  0.475$\pm$0.006 & -0.048$\pm$0.002 &  0.602$\pm$0.001 &  0.045$\pm$0.020 &  0.659$\pm$0.014 \\
{} & {log sSFR} &  0.035$\pm$0.006 & -8.536$\pm$0.003 & -0.007$\pm$0.015 & -8.588$\pm$0.008 &  0.049$\pm$0.002 & -8.730$\pm$0.001 & -0.041$\pm$0.019 & -8.789$\pm$0.013 \\
\enddata
\end{deluxetable*}


\end{document}